\pdfoutput=1

\documentclass[preprintnumbers,showpacs,floats,twocolumn,prd,aps]{revtex4}
\usepackage{amssymb,amsmath}
\usepackage{epsfig,hyperref}
\usepackage[svgnames]{xcolor}
\usepackage{pgf,tikz}

\makeatletter
\renewcommand{\vec}[1]{\mbox{\boldmath$#1$}}
\newcommand{\arXiv}[2][]{\href{http://arxiv.org/abs/#2}{\texttt{arXiv:#2\@ifempty{#1}{}{ [#1]}}}}
\makeatother

\begin{document}

\title{Small Cosmological Constant from\\ Running Gravitational Coupling}
\author{Andrei V. Frolov}\email{frolov@sfu.ca}
\author{Jun-Qi Guo}\email{jga35@sfu.ca}
\affiliation{
  Department of Physics,
  Simon Fraser University\\
  8888 University Drive,
  Burnaby, BC Canada
  V5A 1S6
}
\date{September 3, 2012}

\begin{abstract}
  In this paper, we explore an idea of having Newton's constant change its value depending on the curvature scale involved. Such modification leads to a particular scalar-tensor gravity theory, with the Lagrangian derived from renormalization group (RG) flow arguments. Several of the well-known $f(R)$ modified gravity models have remarkably simple description in terms of the infrared renormalization group, but not the ``designer'' types in general. We find that de Sitter-like accelerated expansion can be generated even in the absence of cosmological constant term, entirely due to running of the Newton's constant. In hopes of tackling the problem of cosmological constant's smallness, we explore the flows which are capable of generating exponential hierarchy between infrared and ultraviolet scales, and investigate cosmological evolution in the models thus derived.
\end{abstract}

\pacs{95.36.+x, 04.50.Kd, 11.10.Hi, 98.80.Cq}
\keywords{}
\preprint{SCG-2011-01}
\maketitle

\section{Introduction}

It has long been hoped that quantum theory of gravity, at least in some limit, allows description in terms of an effective field theory \cite{Weinberg:2009bg, Weinberg:2009wa}. The usual Einstein-Hilbert action is merely the first two terms of an effective action
\begin{equation}\label{eq:eft}
  S[\lambda] =
   \int\left\{\sum\limits_{n=0}^{\infty} \lambda^{4-2n} g^{(n)}(\lambda)\, {\cal R}^{(n)} + \ldots \right\}\sqrt{-g}\,d^4x,
\end{equation}
expanded in local $n$-th order curvature invariants ${\cal R}^{(n)}$, which explicitly depends on the ultraviolet cutoff scale $\lambda$. Should the effective theory of gravity be asymptotically safe, this would allow a sensible ultraviolet-complete description \cite{Niedermaier:2006ns, Percacci:2007sz}. However, recent observations of accelerated expansion of the Universe indicate the presence of a tiny but nonvanishing cosmological constant \cite{Riess:1998cb, Perlmutter:1998np, Riess:2004nr, Komatsu:2010fb}, which presents technical hierarchy problem between (infrared) cosmological acceleration scale and (ultraviolet) Planck scale, which seems irreconcilable despite the many efforts put forward \cite{Weinberg:1988cp, Peebles:2002gy}.

Given enormous separation of cosmological and Planck scales, it would seem highly dubious that quantum gravity is somehow responsible for cosmological acceleration. Corrections and corresponding beta functions reviewed in \cite{Niedermaier:2006ns, Percacci:2007sz} originate from ultraviolet degrees of freedom; they are Planck-suppressed and irrelevant for cosmology. But the language of effective field theory is universal, and should be applicable to infrared phenomena on cosmological scales as well. To give cosmological acceleration a dynamical origin without invoking a new dark energy matter component, one would need a ``modified'' gravity, with new infrared degrees of freedom becoming active at cosmological scales, while remaining hidden in solar system and local tests. A lot has been said on the subject, with $f(R)$ gravity models in particular receiving a lot of attention as of late \cite{Sotiriou:2008rp}, yet the question of large hierarchy ever remains in these attempts.

The main aim of this paper is to apply the idea of renormalization group, which has been very fruitful in high energy physics, to description of these new (and still unknown) infrared degrees of freedom. This obviously entails the shift of attention from ultraviolet cutoff in the effective action (\ref{eq:eft}) to the lower limit of integration -- the {\em infrared cutoff}, and on how things scale when it is varied. Of particular interest is the question whether it is even in principle possible to generate anomalously low scale of cosmological constant by specific running, perhaps in a way similar to dimensional transmutation phenomenon, which is what we will discuss here.

Analyzing renormalization flows dependent on curvature scale, we find that de Sitter-like accelerated expansion can be generated even in the absence of cosmological constant term, entirely due to running of the Newton's constant. This presents a novel way to view the hierarchy problem, in contrast with previous studies \cite{Machado:2007ea, Shapiro:2009dh} involving running of $\Lambda$. Cosmological constant should be set to zero and protected by symmetry, while de Sitter asymptote is attributable to flow of Newton's constant, which we describe by renormalization group equations. Several historically important modified gravity models of $f(R)$ type \cite{Starobinsky:1980te, Capozziello:2003tk, Carroll:2003wy} turn out to have a very simple description in this terms, but not the ``designer'' types, in general.

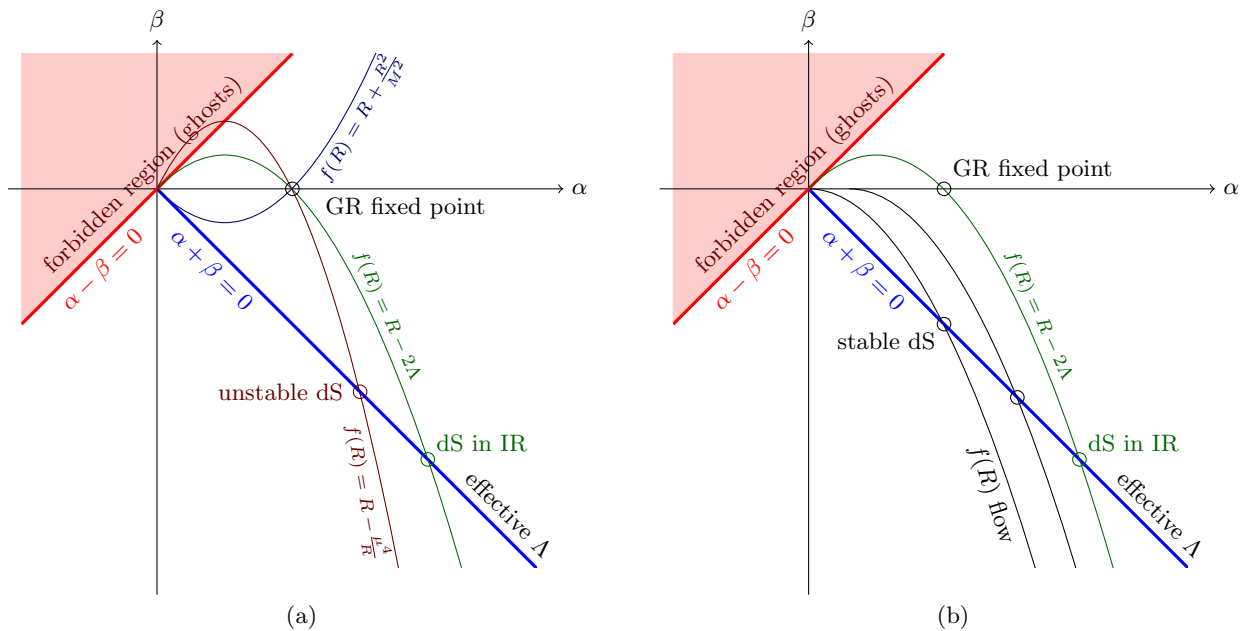
\begin{figure*}
  \begin{center}
  \begin{tabular}{c@{\hspace{24pt}}c}
  \begin{tikzpicture}[scale=1.8]
    \filldraw[color=red!20!white] (-1,-1) -- (1,1) -- (-1,1) -- cycle;
    \node[color=red!40!black] at (-0.1,0.1) {\rotatebox{45}{forbidden region (ghosts)}};
    \node[color=black] at (2.6,-2.4) {\rotatebox{-45}{effective $\Lambda$}};
    \draw[->] (-1.1,0) -- (3,0) node[right] {$\alpha$};
    \draw[->] (0,-3) -- (0,1.1) node[above] {$\beta$};
    \clip (-1,1) rectangle (2.8,-2.8);
    \draw[very thick, color=blue] (0,0) -- (3,-3);
    \draw[very thick, color=red] (-1,-1) -- (1,1);
    \node[color=red] at (-0.4,-0.6) {\rotatebox{45}{$\alpha-\beta=0$}};
    \node[color=blue] at (0.4,-0.6) {\rotatebox{-45}{$\alpha+\beta=0$}};
    \draw[color=green!40!black] (0,0) parabola bend (0.5,0.25) (3,-6);
    \draw[color=blue!40!black] (0,0) parabola bend (0.5,-0.25) (3,6);
    \draw[color=red!40!black] (0,0) parabola bend (0.5,0.50) (3,-12);
    \draw (1,0) circle (0.05) node[below right] {~~~GR fixed point};
    \draw[color=green!40!black] (2,-2) circle (0.05) node[above right] {dS in IR};
    \draw[color=red!40!black] (1.5,-1.5) circle (0.05) node[left] {unstable dS~~};
    \node[above,color=green!40!black] at (1.7,-1.5) {\rotatebox{-66}{\scriptsize$f(R)=R-2\Lambda$}};
    \node[below,color=blue!40!black] at (1.5,1.12) {\rotatebox{60}{\scriptsize$f(R) = R +\! \frac{R^2}{M^2}$}};
    \node[below,color=red!40!black] at (1.55,-1.65) {\rotatebox{-80}{\scriptsize$f(R) = R - \!\frac{\mu^4}{R}$}};
  \end{tikzpicture} &
  \begin{tikzpicture}[scale=1.8]
    \filldraw[color=red!20!white] (-1,-1) -- (1,1) -- (-1,1) -- cycle;
    \node[color=red!40!black] at (-0.1,0.1) {\rotatebox{45}{forbidden region (ghosts)}};
    \node[color=black] at (2.6,-2.4) {\rotatebox{-45}{effective $\Lambda$}};
    \draw[->] (-1.1,0) -- (3,0) node[right] {$\alpha$};
    \draw[->] (0,-3) -- (0,1.1) node[above] {$\beta$};
    \clip (-1,1) rectangle (2.8,-2.8);
    \draw[very thick, color=blue] (0,0) -- (3,-3);
    \draw[very thick, color=red] (-1,-1) -- (1,1);
    \node[color=red] at (-0.4,-0.6) {\rotatebox{45}{$\alpha-\beta=0$}};
    \node[color=blue] at (0.4,-0.6) {\rotatebox{-45}{$\alpha+\beta=0$}};
    \draw (0,0) parabola (3,-9);
    \draw (0.3,0) parabola (3.3,-9);
    \draw[color=black] (1.541619849,-1.541619849) circle (0.05);
    \draw[color=green!40!black] (0,0) parabola bend (0.5,0.25) (3,-6);
    \draw (1,0) circle (0.05) node[above right] {GR fixed point};
    \draw[color=green!40!black] (2,-2) circle (0.05) node[above right] {dS in IR};
    \node[above,color=green!40!black] at (1.7,-1.5) {\rotatebox{-66}{\scriptsize$f(R)=R-2\Lambda$}};
    \draw[color=black] (1,-1) circle (0.05) node[below left] {stable dS};
    \node[below] at (1.35,-1.8) {\rotatebox{-72}{$f(R)$ flow}};
  \end{tikzpicture} \\
  (a) & (b) \\
  \end{tabular}
  \end{center}
  \caption{Renormalization group flows causing cosmological acceleration: several known $f(R)$ models that can be generated by renormalization group (a), and possible flows generating exponential hierarchy between infrared and ultraviolet scales (b).}
  \label{fig:flow}
\end{figure*}

We argue that a ``soft'' running, in which Newton's constant flows with quadratic beta function approaching a high-curvature fixed point, generates exponential hierarchy between infrared and ultraviolet scales, which could be exploited for cosmological model-building. The simplest model of this type, with beta function analogous to the one for QCD coupling constant, is somewhat problematic observationally (due to Newton's constant running to zero in high curvature limit), but is easy to analyze analytically. We present it here due to its simplicity, and then discuss more realistic models with exponential hierarchy derived by this method, which did not appear in the literature before.

\section{Infrared Renormalization Group and Modified Gravity}

Let's explore the implications of a running Newton's constant with an energy scale set by curvature scalar $R$, which we will describe by introducing a dimensionless coupling constant $\alpha$ by
\begin{equation}\label{eq:alpha}
  8\pi G = \alpha\, m_{\text{pl}}^{-2}.
\end{equation}
Note that presence of reduced Planck mass $m_{\text{pl}}$ in the above definition means that gravitational coupling is compared to a fixed scale (namely the ultraviolet limit value), which only makes sense if one is talking about its change with infrared cutoff, as we do here.
We will assume that explicit cosmological constant term in effective action (\ref{eq:eft}) strictly vanishes, and we will ignore higher order curvature corrections as we are primarily interested in infrared effects. If renormalization group flow is autonomous, the running of the dimensionless coupling with scale $\mu \equiv R/R_0$ is described by a beta-function flow
\begin{equation}\label{eq:beta}
  \mu\, \frac{d\alpha}{d\mu} = \beta(\alpha),
\end{equation}
which is integrable for one-dimensional dynamical system
\begin{equation}
  \int \frac{d\alpha}{\beta(\alpha)} = \int \frac{d\mu}{\mu}.
\end{equation}
Thus, Newton's constant in Einstein-Hilbert action becomes a function of curvature, and the theory of gravity is promoted to $f(R)$-type \cite{Sotiriou:2008rp} with Lagrangian
\begin{equation}\label{eq:L}
  {\cal L}_{\text{GR}} = \frac{R}{16\pi G} \mapsto
    \frac{m_{\text{pl}}^{2}}{2}\, \frac{R}{\alpha} = {\cal L}_{f(R)}.
\end{equation}
The crucial point here is that autonomous flow (\ref{eq:beta}) is independent of $R_0$ (which is merely a reference scale), and $f(R) \equiv R/\alpha$ generated by it does not explicitly involve a tiny dimensionful scale $R_{\text{IR}}$ introduced by hand in most if not all $f(R)$ models proposed so far \cite{Capozziello:2003tk, Carroll:2003wy, Hu:2007nk, Starobinsky:2007hu, Appleby:2007vb, Miranda:2009rs}, which replaces cosmological constant but of course still presents the same hierarchy problem. Instead, the hierarchy appears in {\em particular solutions} connecting local high-curvature environment (where we seem to live right on top of ultraviolet fixed point) and the cosmological acceleration attractor in the infrared limit. We will return to this point again in the next section, but for now let's explore $f(R)$ models generated by such flows.

For $f(R)$ gravity to be perturbatively stable, one needs $f'>0$ to avoid ghosts \cite{Nunez:2004ji}, and $f''>0$ to avoid tachyon instability \cite{Dolgov:2003px, Sawicki:2007tf}. These conditions are readily rewritten in terms of flow equation quantities
\begin{equation}\label{eq:derivs}
  f' = \frac{\alpha-\beta}{\alpha^2}, \hspace{1em}
  R f'' = - \left(1 - 2\, \frac{\beta}{\alpha} + \beta'\right) \frac{\beta}{\alpha^2}.
\end{equation}
In addition, $f(R)$ gravity equations of motion (\ref{eq:eom},\ref{eq:trace}) allow a vacuum (anti-) de Sitter solution if
\begin{equation}
  2f-f'R = \frac{\alpha+\beta}{\alpha^2}\, R = 0.
\end{equation}
From (\ref{eq:derivs}) it follows that de Sitter solution is unstable if $\beta' < -3$. If one wishes to have asymptotically safe theory with effective cosmological constant, the renormalization group flow should connect $\alpha+\beta=0$ line in IR to a $\beta=0$ fixed point in UV, while avoiding forbidden region $\beta>\alpha$. A number of such flows is shown in both panels of Fig.~\ref{fig:flow}, both with stable and unstable de Sitter vacua.

Interestingly enough, Einstein-Hilbert action with cosmological constant, Einstein-Hilbert action with ultraviolet $R^2$ corrections \cite{Starobinsky:1980te}, and original $1/R$ proposal for infrared corrections \cite{Capozziello:2003tk, Carroll:2003wy} are all generated by very simple renormalization flows, as shown in Fig.~\ref{fig:flow}a, all of which, jumping ahead of ourselves, only differ by a single coefficient of proportionality (\ref{eq:beta:power}). For example, requiring $f'=1$ identically in (\ref{eq:derivs}), we should recover general relativity action. Indeed, integrating flow (\ref{eq:beta}) with 
\begin{equation}\label{eq:beta:gr}
  \beta(\alpha) = \alpha - \alpha^2
\end{equation}
leads to a rational dependence of $\alpha$ on curvature
\begin{equation}
  \alpha = \frac{\mu}{\mu-1},
\end{equation}
which puts back the constant term in the action
\begin{equation}\label{eq:f:gr}
  f(R) \equiv \frac{R}{\alpha} = R - R_0.
\end{equation}
All power-law corrections to Einstein-Hilbert action
\begin{equation}\label{eq:f:power}
  f(R) = R \left[1 + \lambda \left(\frac{R}{R_0}\right)^{\!\!n}\, \right]
\end{equation}
are in fact generated by an autonomous flow
\begin{equation}\label{eq:beta:power}
  \beta(\alpha) = n\, \alpha(\alpha-1).
\end{equation}
This could be checked by direct integration as we did above, but it is instructive to ``reverse engineer'' the flow (\ref{eq:beta:power}) from expression (\ref{eq:f:power}), to see how the scale $R_0$ disappears from the flow equations. Writing $\alpha \equiv R/f(R)$
\begin{equation}\label{eq:alpha:power}
  \alpha = \frac{1}{1+\lambda\mu^n},
\end{equation}
and formally taking logarithmic derivative $\beta = \mu\, d\alpha/d\mu$, we obtain the expression for $\beta$ as a function of scale $\mu$
\begin{equation}
  \beta = - \frac{n\lambda\mu^n}{(1+\lambda\mu^n)^2}.
\end{equation}
This by itself is simply an identical rewrite of expressions for $f(R)$ and its derivative $f'$ in new variables, but if scale $\mu$ could be eliminated in favour of the coupling value $\alpha$, the dynamical system would close and become autonomous. Inverting equation (\ref{eq:alpha:power}), we see that $\beta$ indeed can be written as a function of $\alpha$ only
\begin{equation}
  \beta = -n\alpha^2(\alpha^{-1}-1),
\end{equation}
which leads to the autonomous flow (\ref{eq:beta:power}).

\begin{figure}
  \centerline{\epsfig{file=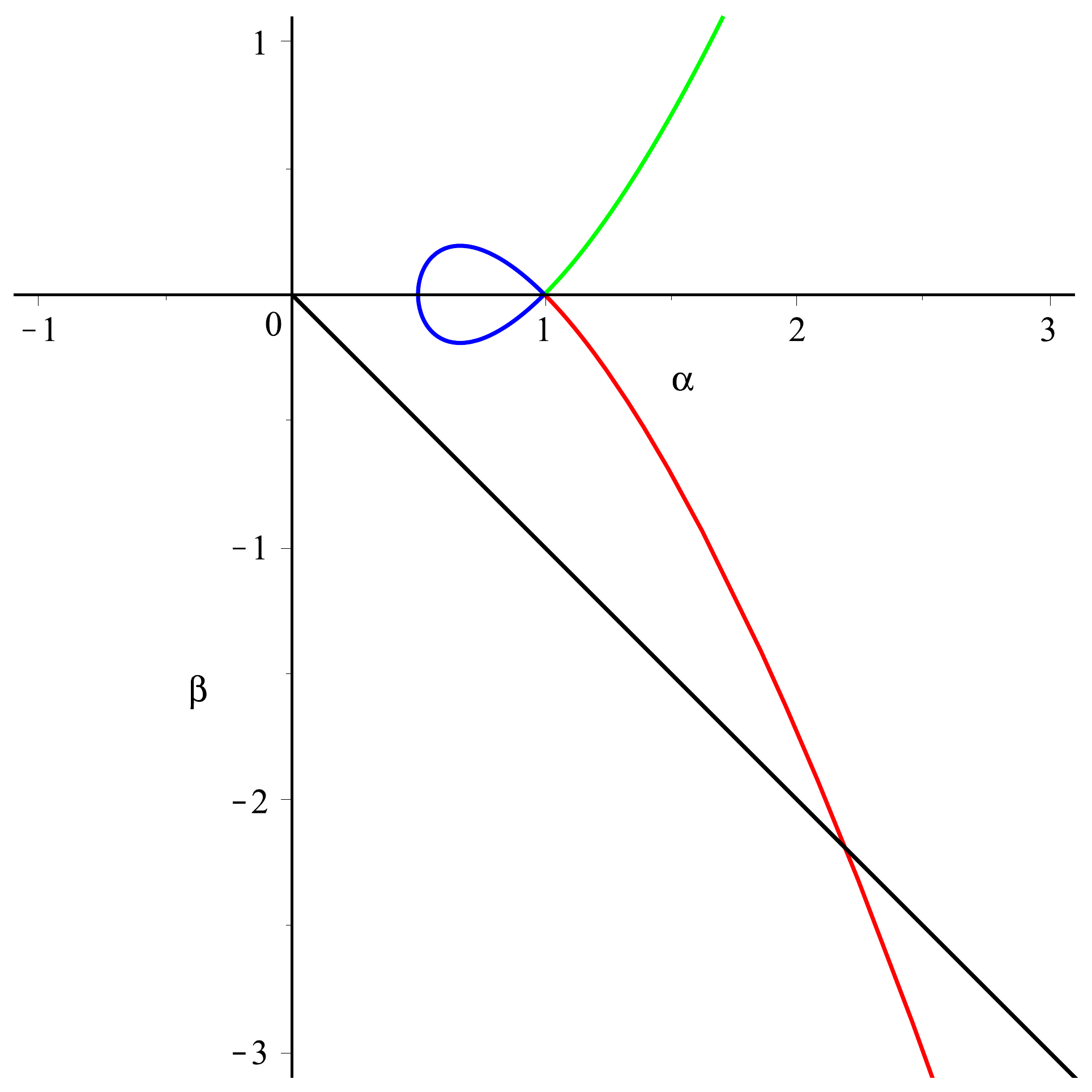, width=240pt}}
  \caption{Not all $f(R)$ models have a simple renormalization group flow interpretation -- e.g.\ Starobinsky model \cite{Starobinsky:2007hu} shown.}
  \label{fig:starobinsky}
\end{figure}

Even in one-dimensional dynamical systems of the type we consider, this inversion is not guaranteed to happen. Thus, not all $f(R)$ models allow simple interpretation in terms of renormalization group flow. As a representative example, let's consider Starobinsky model \cite{Starobinsky:2007hu}
\begin{equation}\label{eq:f:starobinsky}
  f(R) = R \left[ 1 + \frac{\lambda}{\mu} \Big( (1+\mu^2)^{-n} - 1\Big) \right], \hspace{1em}
  \mu \equiv \frac{R}{R_0}.
\end{equation}
This function corresponds to a non-invertible relationship between coupling strength $\alpha$ and curvature scale $\mu$
\begin{equation}
  \alpha^{-1} \equiv \frac{f(R)}{R} = 1 + \frac{\lambda}{\mu} \Big( (1+\mu^2)^{-n} - 1\Big),
\end{equation}
which leads to a multi-valued function $\beta(\alpha)$ as shown in Fig.~\ref{fig:starobinsky} (for particular values of $n=1$ and $\lambda=2$). Following the flow along cosmologically relevant branch (shown in red) from de Sitter solution toward increasing curvature, one jumps through $R \rightarrow \pm\infty$ at $\alpha=1$, switching over to the negative curvature branch (shown in blue), and then passing through $R=0$ at $\alpha=1$ again to finally reach small positive curvature branch (shown in green).

\section{Implications for Hierarchy Problem}

General behavior of modified gravity theories described by renormalization group flow (\ref{eq:beta}) in high-curvature limit is determined by an asymptotic expansion of the flow near the ultraviolet fixed point $\alpha_*$
\begin{equation}\label{eq:beta:UV}
  \beta(\alpha) \approx \sum\limits_{k=1}^{\infty} c_k (\alpha-\alpha_*)^k.
\end{equation}
In the models we discussed so far, this series starts from a linear term (as shown in Fig.~\ref{fig:flow}a), and integrating the flow leads to a power-law relationships between scale $\mu$, coupling $\alpha$, and running $\beta$. With respect to cosmological curvature scale, there is little difference between local environment on Earth and Planck scale; both are infinitesimally close to UV fixed point, and there are no obvious reasons why the scale separation should be so large.

But if approach to UV fixed point $\alpha_*$ is slower than linear (i.e.\ $c_1=0$) instead (as shown in Fig.~\ref{fig:flow}b), integration of flow equations leads to a logarithmic dependence of coupling $\alpha$ and running $\beta$ on the curvature scale $\mu$, which means that reasonable separations of the points on the flow diagram $\beta(\alpha)$ can become exponentially large in physical curvature. Placing an upper limit on the running $\beta$ of Newton's constant in near-Earth environment (for example, by null deviation from Newton's force law in torsional pendulum experiments \cite{Adelberger:2009zz}), we then suddenly have something to say about how far we should be removed from the infrared scale so as not to detect any running.

\begin{figure}
  \centerline{\epsfig{file=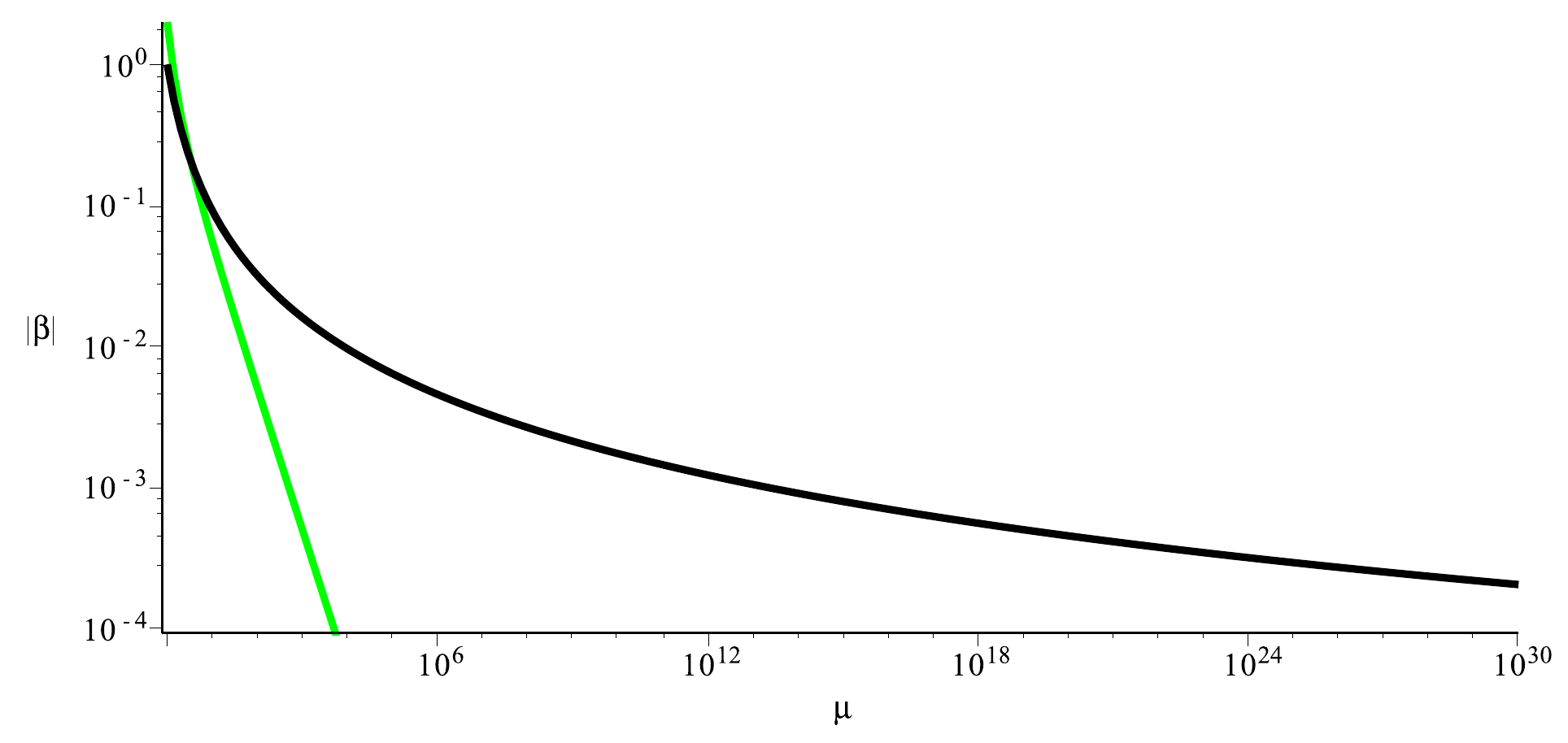, width=240pt}}
  \caption{Running $\beta$ as a function of scale $\mu \equiv R/R_0$ for linear (green) and quadratic (black) approach to UV fixed point.}
  \label{fig:bound}
\end{figure}

This argument is illustrated by Fig.~\ref{fig:bound},
which shows running $\beta \equiv d\alpha/d\ln\mu$ as a function of curvature scale $\mu$ for two flows, $\beta = -\alpha(\alpha-1)$ (in green) and $\beta = -\alpha^2$ (in black). For example, placing an (arbitrary) upper limit on $|\beta| < 10^{-4}$ does not tell us much about the first flow, yet requires a separation of at least 30 orders of magnitude in curvature between infrared and measurement scales for the second flow. In other words, deviations from Einstein gravity depend on how close we are to UV fixed point in the lab measurements, thus the smallness of cosmological constant is tied down to the observation that $f(R)$ gravity reproduces Einstein theory!

This simple thought could be used to generate small numbers involved in the hierarchy problem in cosmological model-building; the only thing we need to change is to make beta function quadratic in UV. To illustrate the general idea, let's consider the simplest model of such kind, namely the one where Newton's constant runs to zero with quadratic beta function in UV limit (which is similar to running of the QCD coupling constant)
\begin{equation}\label{eq:beta:IR}
  \beta(\alpha) = - \alpha^2.
\end{equation}
Integrating the flow equation (\ref{eq:beta:IR}) leads to an effective action for gravity with
\begin{equation}\label{eq:f}
  f(R) = \frac{R}{\alpha_0} \left(1 + \alpha_0 \ln\frac{R}{R_0}\right).
\end{equation}
Here $\alpha_0$ is the coupling measured at curvature scale $R_0$, and its value also determines how fast the Newton's constant runs at that scale. Note that $R_0$ is not really a parameter in the action, but merely the scale at which the action is evaluated. Coupling $\alpha$ can be renormalized to another arbitrary scale $R_1$ by redefining
\begin{equation}
  \alpha_1 = \frac{\alpha_0}{1+\alpha_0\ln\frac{R_1}{R_0}},\\
\end{equation}
without any change to the Lagrangian itself. Corrections of the form (\ref{eq:f}) already appeared in \cite{Amendola:2006we, Kainulainen:2008pr}, although full implications for hierarchy were not realized then.

The obvious problem with the simplistic model (\ref{eq:beta:IR}) is that Newton's constant {\em actually does run} to zero, which of course is kind of different from the expected behaviour in large curvature limit. This drawback is easily rectified if one considers Newton's constant running to a constant
\begin{equation}\label{eq:beta:2}
  \beta(\alpha) = - \kappa (\alpha-\alpha_*)^2.
\end{equation}
Integrating this flow leads to a rational function of logarithm of curvature of the form
\begin{equation}\label{eq:f:2}
  f(R) = \frac{a + b\,\ln\mu}{c + d\,\ln\mu}\, R, \hspace{1em}
  \mu \equiv \frac{R}{R_0}.
\end{equation}
As before, constants in the action can be redefined by changing the reference scale $R_0$, but the group action is slightly more complicated. Running of $\beta$ with curvature scale $\mu$ is essentially the same as in simpler model (\ref{eq:beta:IR}), and is not shown in Fig.~\ref{fig:bound} to avoid crowding the plot. While more realistic, this model is more cumbersome to analyze analytically. As far as we know, it has not been discussed before.

\section{Cosmological Evolution}

Now let's discuss cosmological evolution of the models we introduced in the previous section. Variation of the Lagrangian (\ref{eq:L}) with respect to the metric $g_{\mu\nu}$ yields field equations of motion in $f(R)$ gravity
\begin{equation}\label{eq:eom}
  f' G_{\mu\nu} - f'_{;\mu\nu} + \left[\Box f'  - \frac{1}{2}\, (f-f'R)\right] g_{\mu\nu} = m_{\text{pl}}^{-2}\, T_{\mu\nu},\!
\end{equation}
which involve higher derivatives of the metric, as usual in $f(R)$ gravity. Taking a trace of the above equation
\begin{equation}\label{eq:trace}
  \Box f' = \frac{1}{3}\, (2f-f'R) + m_{\text{pl}}^{-2}\, \frac{T}{3},
\end{equation}
and introducing a scalar degree of freedom $\phi \equiv f'-2$, equations (\ref{eq:eom},\ref{eq:trace}) are cast into a scalar-tensor theory form
\begin{equation}\label{eq:st}
  G_{\mu\nu} = m_{\text{pl}}^{-2}\, T_{\mu\nu} + {\cal Q}_{\mu\nu}, \hspace{1em}
  \Box\phi = V'(\phi) - {\cal F}.
\end{equation}
Here ${\cal Q}_{\mu\nu}$ is the effective stress-energy tensor of geometric degree of freedom $\phi$ (sometimes referred to as {\em scalaron})
\begin{equation}\label{eq:Q}
  {\cal Q}_{\mu\nu} = -(1+\phi)\, G_{\mu\nu} + \phi_{;\mu\nu} - \Big[\Box\phi - 3 P(\phi)\Big]\, g_{\mu\nu},
\end{equation}
with $V'(\phi)$ and $P(\phi)$ defined by equations (\ref{eq:P},\ref{eq:V'}), and ${\cal F} \equiv - m_{\text{pl}}^{-2}\, T/3$ is a matter force term for field $\phi$. As Einstein tensor $G_{\mu\nu}$ can be ``traded'' between left and right sides of equation, definition of ${\cal Q}_{\mu\nu}$ is not unique.

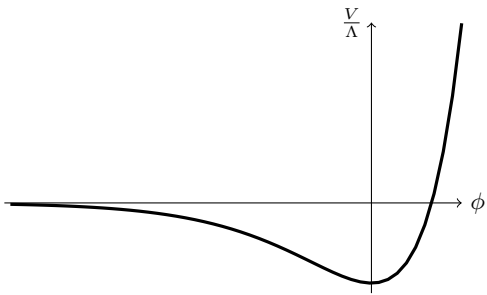
\begin{figure}
  \begin{center}
  \begin{tikzpicture}[scale=0.8]
    \draw[very thick] plot[id=V,domain=-6:1.5,samples=50] function{4*(x-1)*exp(x)/3};
    \draw[->] (-6.1,0) -- (1.5,0) node[right] {$\phi$};
    \draw[->] (0,-1.5) -- (0,3) node[left] {$\frac{V}{\Lambda}$};
  \end{tikzpicture}
  \end{center}
  \caption{Effective potential for scalar field $\phi$.}
  \label{fig:V}
\end{figure}

The above applies for a general $f(R)$ gravity theory. Equations for our model (\ref{eq:f}) become quite simple. Scalar degree of freedom $\phi$ is just a logarithm of the curvature
\begin{equation}\label{eq:phi}
  \phi \equiv f'-2 = \ln\frac{R}{R_0} + \frac{1-\alpha_0}{\alpha_0} = \ln\frac{R}{R_*},
\end{equation}
where the constant in the definition is chosen so that $\phi_* = 0$ solution corresponds to a de Sitter vacuum with
\begin{equation}\label{eq:lambda}
  R_* \equiv 4 \Lambda = R_0 \exp\left[\frac{\alpha_0-1}{\alpha_0}\right].
\end{equation}
Note that if the running parameter is small ($\alpha_0 \ll 1$), effective cosmological constant $\Lambda$ is {\em exponentially suppressed} compared to the reference scale $R_0$! In terms of $\Lambda$, the $\phi$-dependent terms in equations of motion (\ref{eq:st},\ref{eq:Q}) become
\begin{equation}\label{eq:P}
  P(\phi) \equiv \frac{1}{6}\, (f-f'R) =
    -\frac{2}{3}\, \Lambda e^\phi,
\end{equation}
\vspace{-4ex}
\begin{equation}\label{eq:V'}
  V'(\phi) \equiv \frac{1}{3}\, (2f-f'R) =
    \frac{4}{3}\, \Lambda e^\phi \phi.
\end{equation}
Effective potential $V(\phi)$ of the scalar degree of freedom is readily integrated and has a simple analytic form
\begin{equation}\label{eq:V}
  V(\phi) = \frac{4}{3}\, \Lambda e^\phi (\phi-1).
\end{equation}
The potential has a single de Sitter minimum at $\phi=0$, flat small curvature asymptotic ($\phi\rightarrow -\infty$ as $R=0$), an exponential potential wall for large curvature ($\phi\rightarrow +\infty$), 
as plotted in Figure~\ref{fig:V}.
The matter force term drives the equilibrium field value $V'(\phi_{\text{eq}}) = {\cal F}$ up the potential wall
\begin{equation}\label{eq:eq}
  \phi_{\text{eq}} = W\left(\frac{3}{4}\frac{\cal F}{\Lambda}\right),
\end{equation}
with explicit equilibrium position given in terms of Lambert $W$-function. The field $\phi$ is light near de Sitter minimum, but for matter densities substantially higher than effective cosmological constant ${\cal F}/\Lambda\gg 1$, field $\phi$ becomes heavy, and one expects the usual chameleon mechanism to ``freeze'' the scalar degree of freedom \cite{Khoury:2003rn, Khoury:2003aq}.

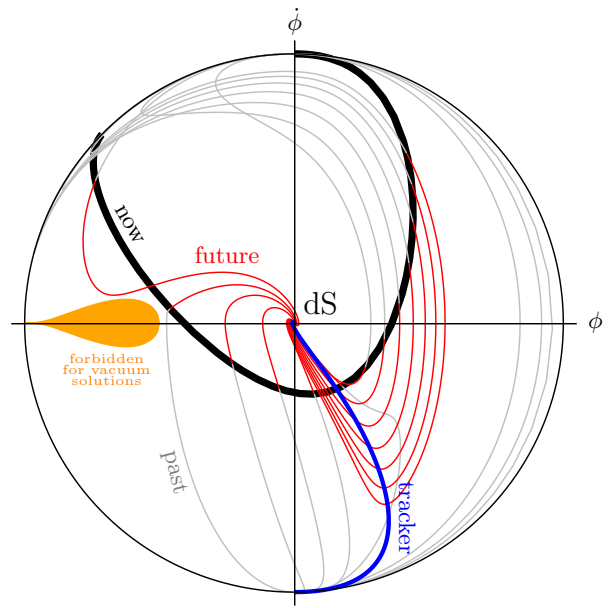
\begin{figure}
  \begin{center}
  \begin{tikzpicture}
    \node at (0,0) {\epsfig{file=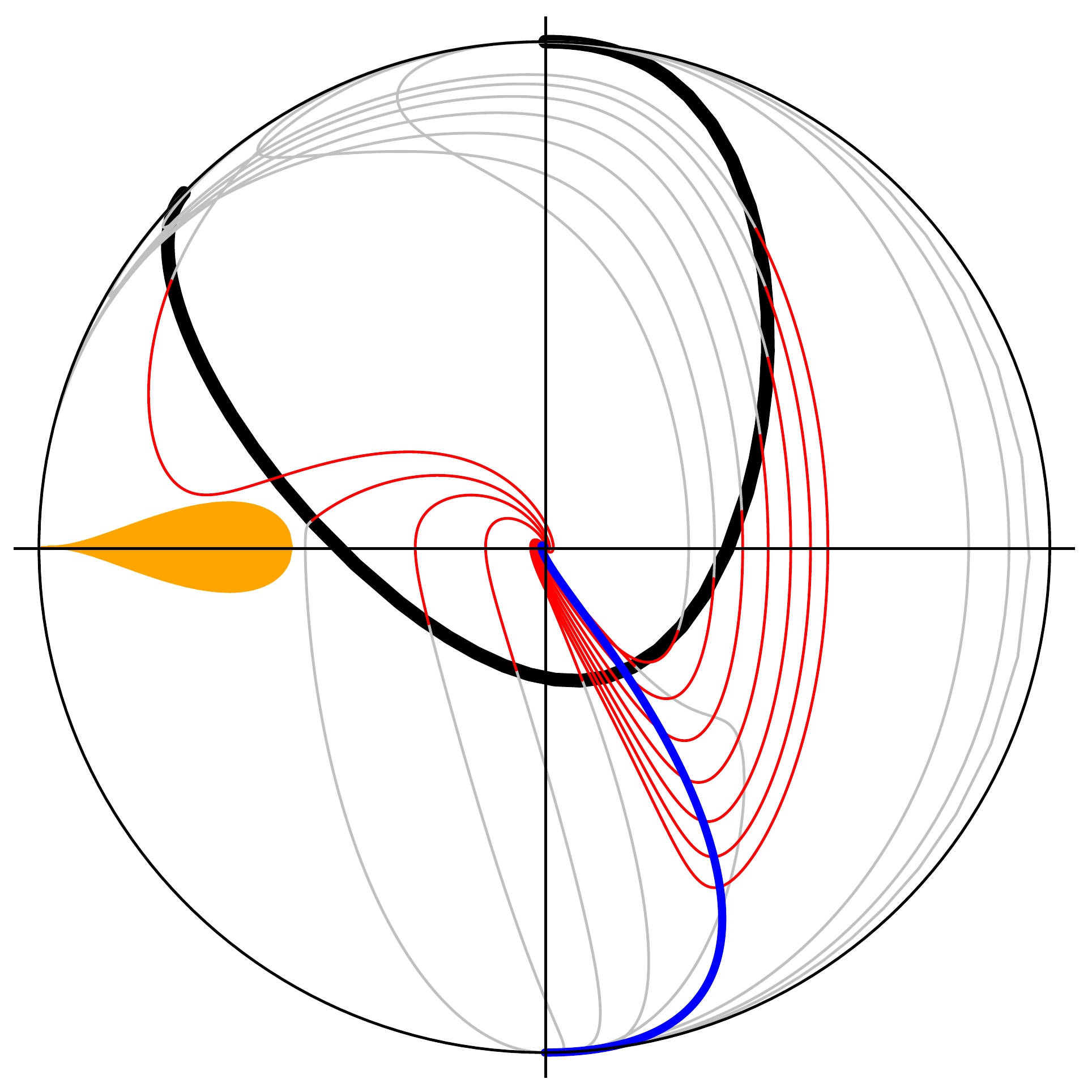, width=220pt}};
    \node at (4,0) {$\phi$};
    \node at (0,4) {$\dot\phi$};
    \node[above right] at (0,0) {\large dS};
    \node[below,color=orange] at (-2.5,-0.3) {\tiny forbidden};
    \node[below,color=orange] at (-2.5,-0.45) {\tiny for vacuum};
    \node[below,color=orange] at (-2.5,-0.6) {\tiny solutions};
    \node at (-2.2,1.3) {\rotatebox{-60}{now}};
    \node[color=red] at (-0.9,0.9) {\rotatebox{0}{future}};
    \node[color=gray] at (-1.6,-2) {\rotatebox{-76}{past}};
    \node[color=blue] at (1.5,-2.6) {\rotatebox{-90}{tracker}};
  \end{tikzpicture}
  \end{center}
  \caption{Poincar\'e projection of dynamical system (\ref{eq:ds}).}
  \label{fig:phase}
\end{figure}

Cosmological evolution of a flat Friedmann-Robertson-Walker universe with a metric
\begin{equation}\label{eq:frw}
  ds^2 = -dt^2 + a^2(t)\, d\vec{x}^2
\end{equation}
in our model can be completely analyzed using dynamical systems techniques. The phase space is four-dimensional, with variables $\{\phi,\pi,a,H\}$ and dynamical equations 
\begin{equation}\label{eq:ds}
  \dot{\pi} + 3H\pi + V'(\phi) = m_{\text{pl}}^{-2}\, \frac{\rho-3p}{3},
  \hspace{1em}
  \dot{H} = \frac{R}{6} - 2H^2,
\end{equation}
supplemented with definitions $\dot\phi=\pi$ and $\dot{a} = aH$.
The evolution is subject to a constraint (which is analogous to Friedmann equation in the usual cosmology)
\begin{equation}\label{eq:constraint}
  H^2(\phi+2) + H\pi + P(\phi) = m_{\text{pl}}^{-2}\, \frac{\rho}{3},
\end{equation}
and hence is restricted to a three-dimensional surface in a phase space (which has a rather complicated shape). The clearest view of the trajectories is allowed by projection on $\vec{x} \equiv \{\phi,\pi\}$ plane followed by Poincar\'e compactification $\vec{y} \equiv \vec{x}/(\sigma^2+\vec{x}^2)^{\frac{1}{2}}$, as shown in Figure~\ref{fig:phase}.
Detailed analysis of the dynamics will be presented elsewhere, here we will just highlight the main features. The only finite critical point of dynamical system (\ref{eq:ds}) is de Sitter attractor, which all cosmological solutions approach in the future. Among all trajectories, the most important is the one tracking the diluting matter energy density
\begin{equation}\label{eq:rho}
  m_{\text{pl}}^{-2}\,\frac{\rho}{3} = \left(\frac{\Omega_{M,0}}{a^3} + \frac{\Omega_{R,0}}{a^4}\right) H_0^2,
\end{equation}
shown in blue in Figure~\ref{fig:phase}. In the past, it closely follows the minimum of the effective potential (\ref{eq:eq}) due to matter
\begin{equation}\label{eq:trac}
  \phi_{\text{trac}} \simeq W\left(\frac{3H_0^2}{4\Lambda}\,\frac{\Omega_{M,0}}{a^3}\right),
\end{equation}
until the scalar field $\phi$ becomes light and ``releases'', approaching de Sitter minimum kinematically. Tracing this trajectory to the present day value of constraint equation (\ref{eq:constraint}), shown as thick black line in Figure~\ref{fig:phase}, gives a cosmological constraint on values of $\Omega_M$, $\Omega_R$ and $\Lambda$, similar to closure relationship in flat $\Lambda$CDM model.
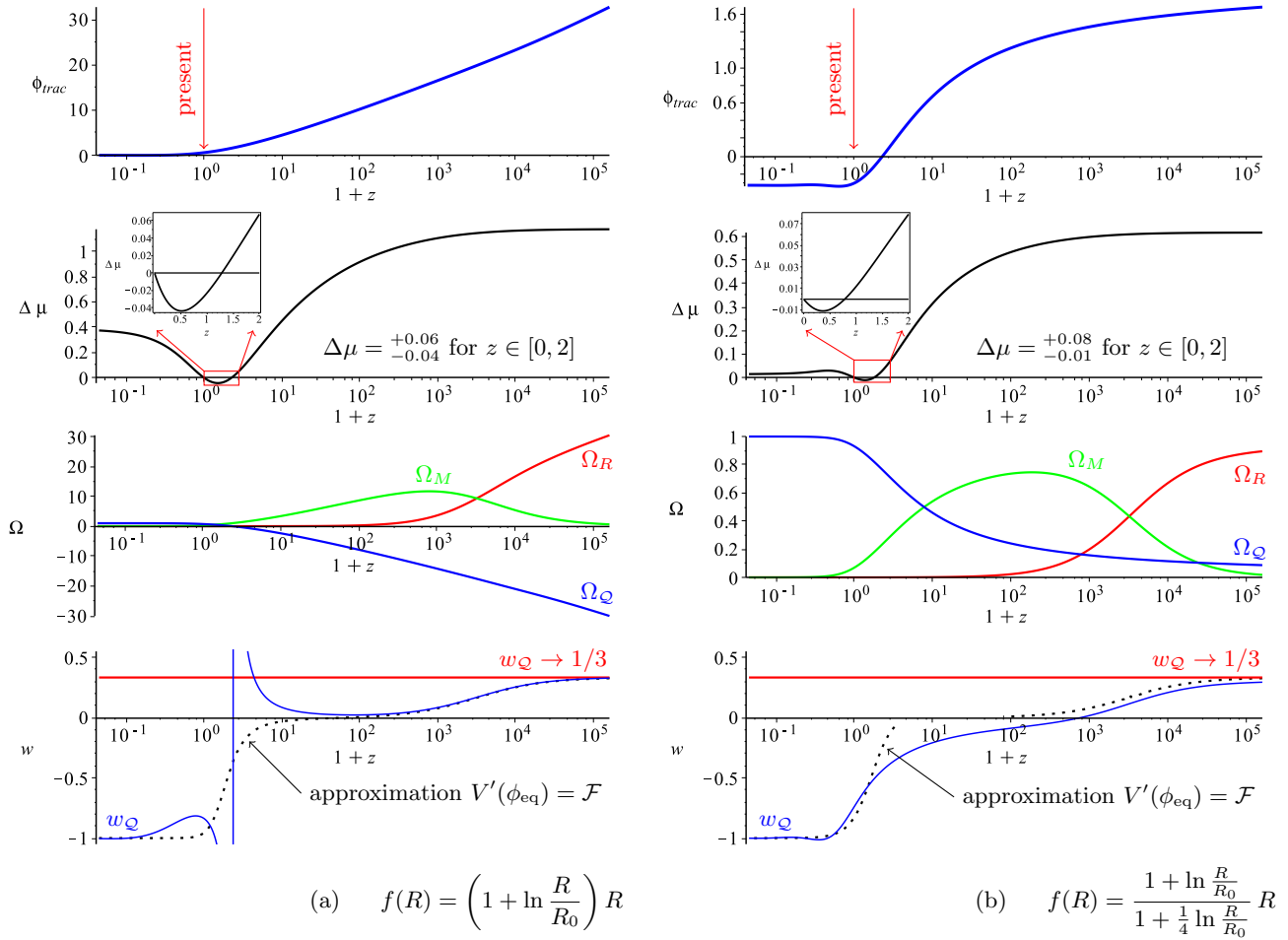
\begin{figure*}
  \begin{center}\begin{tabular}{rr}
  \begin{tikzpicture}
    \node at (0,8.7) {\epsfig{file=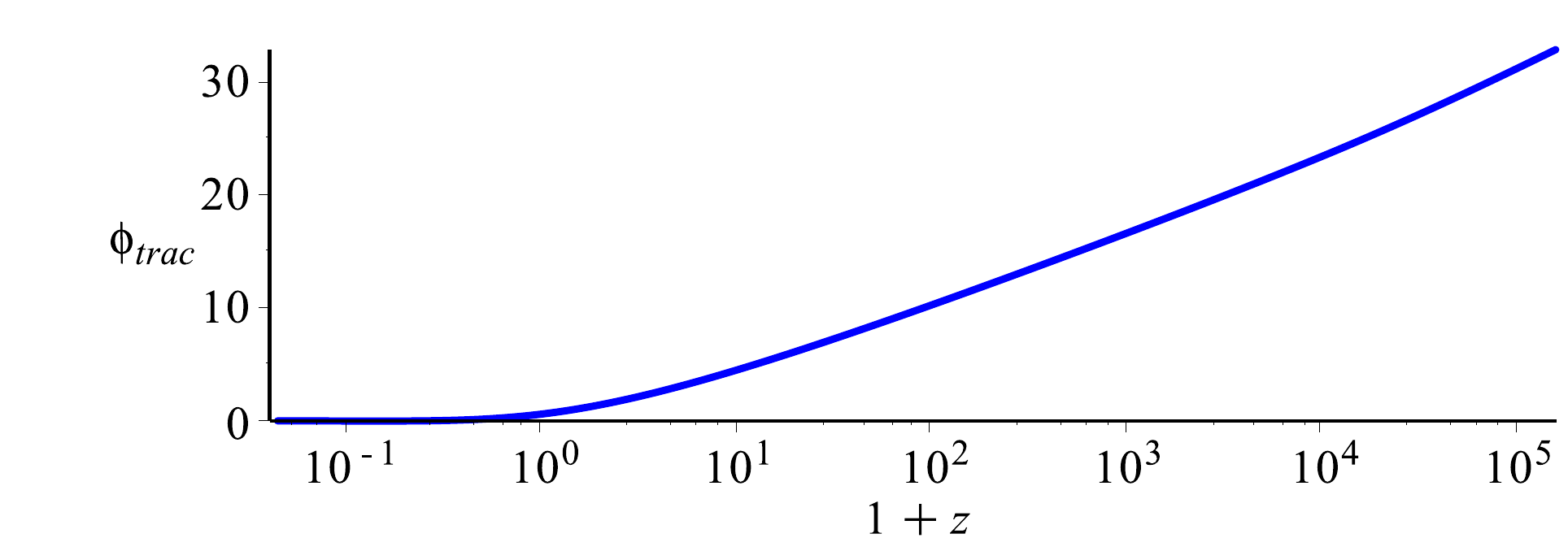,  width=240pt}};
      \draw[->,color=red] (-1.313,9.9) -- (-1.313,8.0) node[midway,left] {\rotatebox{90}{present}};
    \node at (0,5.7) {\epsfig{file=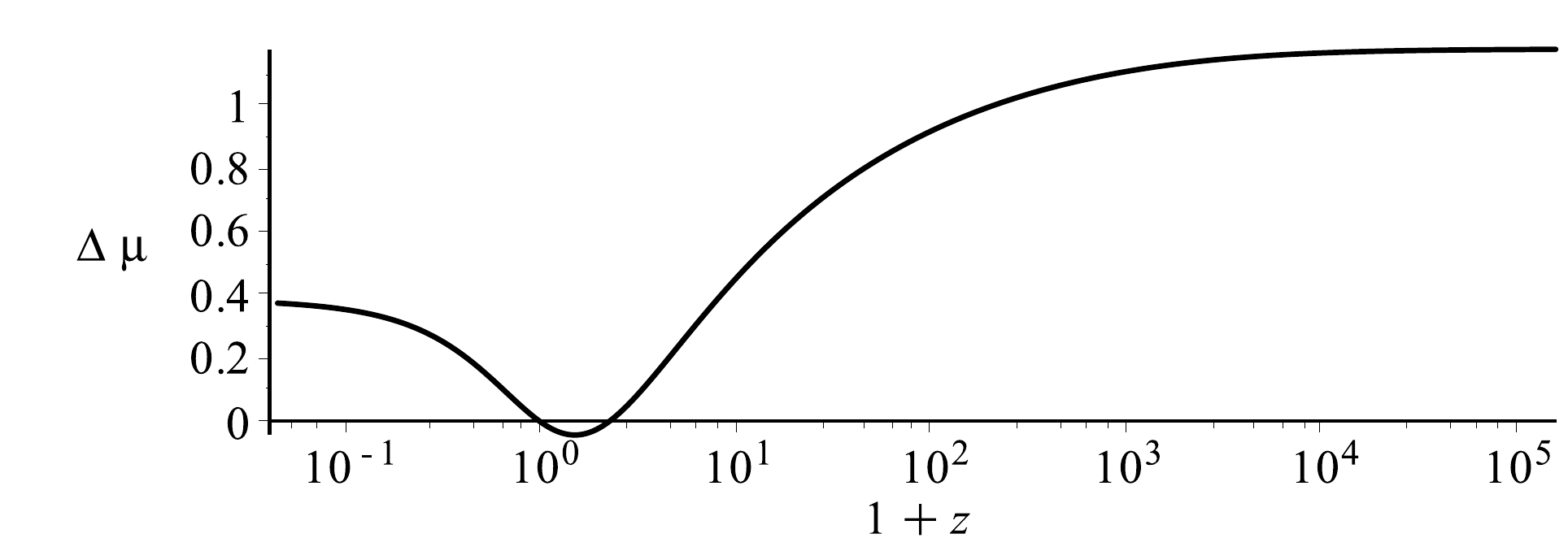, width=240pt}};
    \node at (-1.6,6.3) {\epsfig{file=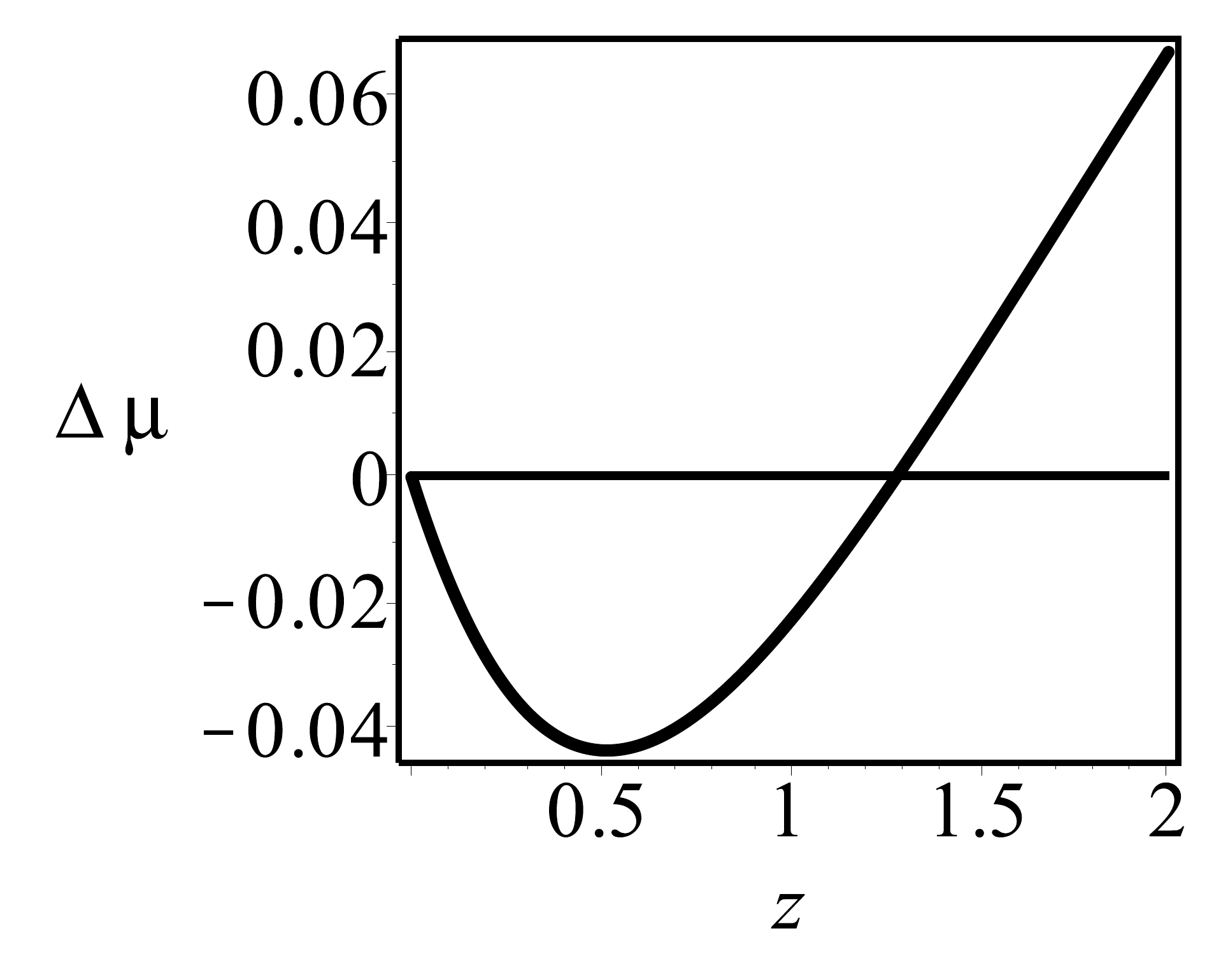, height=52pt}};
      \draw[<-,color=red] (-1.95,5.55) -- (-1.31,5.0);
      \draw[<-,color=red] (-0.65,5.55) -- (-0.84,5.0);
      \draw[color=red] (-1.31,4.81) rectangle (-0.84,5.0);
      \node[above] at (2,5.0) {$\Delta\mu = {}^{+0.06}_{-0.04}$ for $z \in [0,2]$};
    \node at (0,3.0) {\epsfig{file=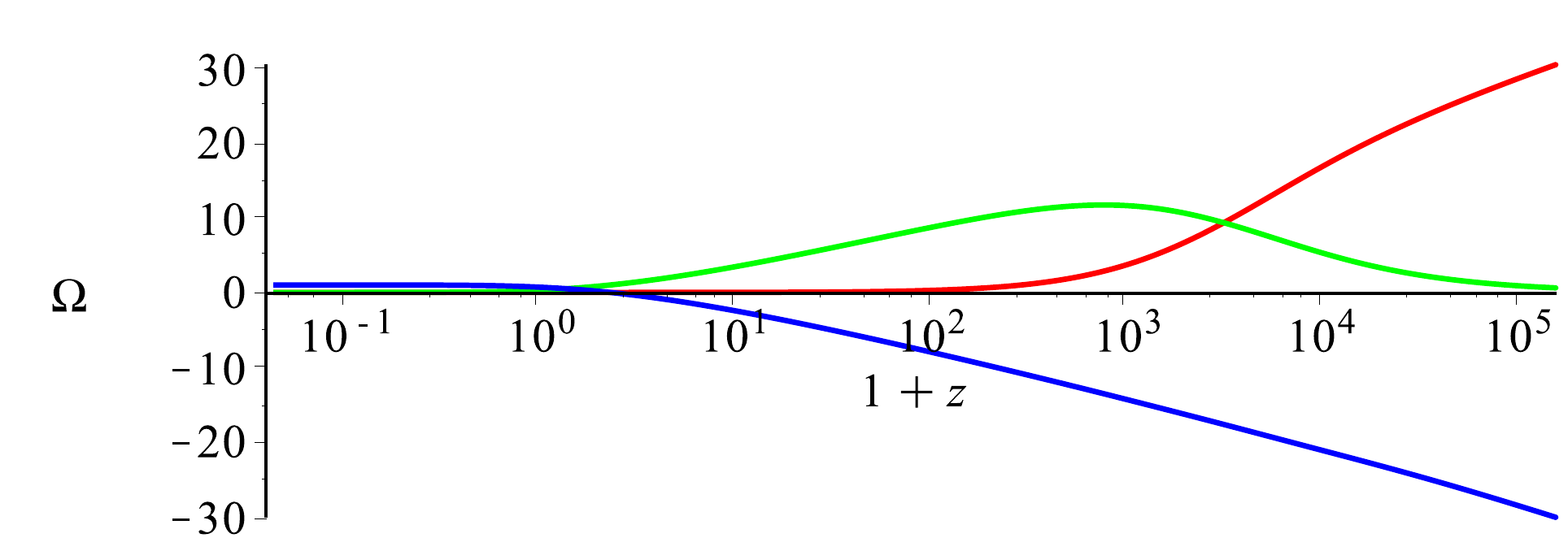, width=240pt}};
      \node[color=red] at (4,3.8) {$\Omega_R$};
      \node[color=green] at (1.8,3.6) {$\Omega_M$};
      \node[color=blue] at (4,2.0) {$\Omega_{\cal Q}$};
    \node at (0,0.0) {\epsfig{file=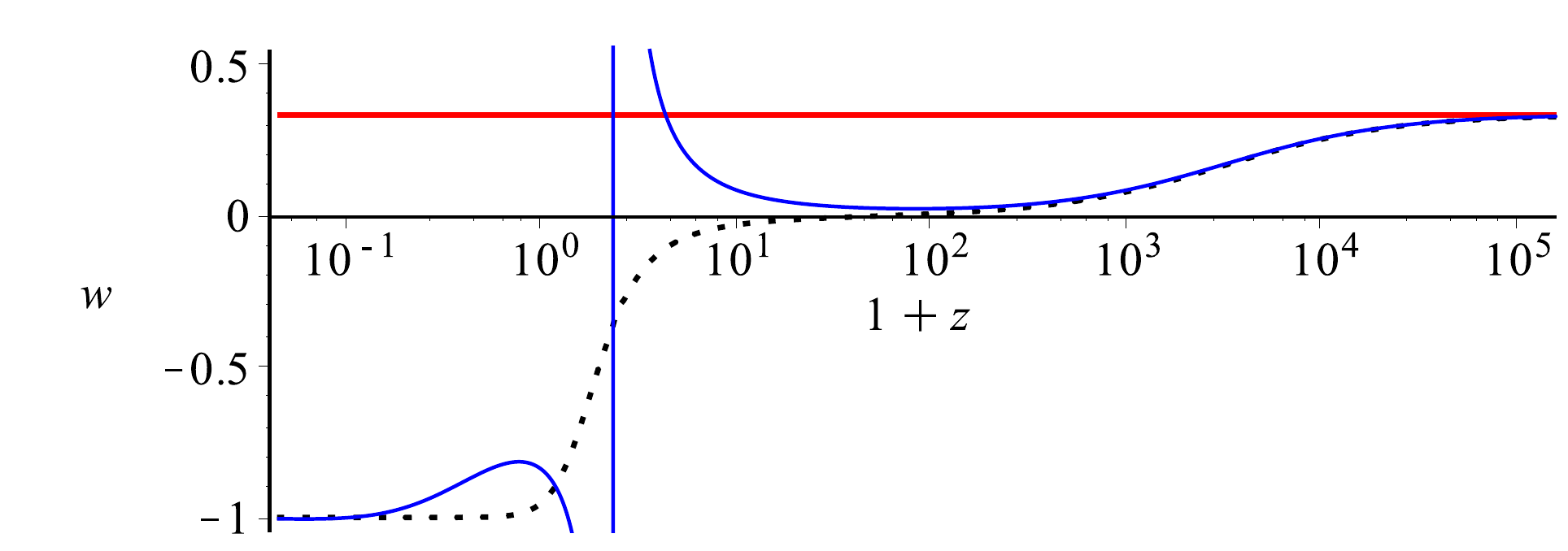, width=240pt}};
      \node[color=blue] at (-2.4,-1.1) {$w_{\cal Q}$};
      \node[color=red] at (3.4,1.1) {$w_{\cal Q} \rightarrow 1/3$};
      \draw[<-] (-0.7,0.0) -- (0,-0.7) node[right] {approximation $V'(\phi_{\text{eq}}) = {\cal F}$};
  \end{tikzpicture} &
  \begin{tikzpicture}
    \node at (0,8.7) {\epsfig{file=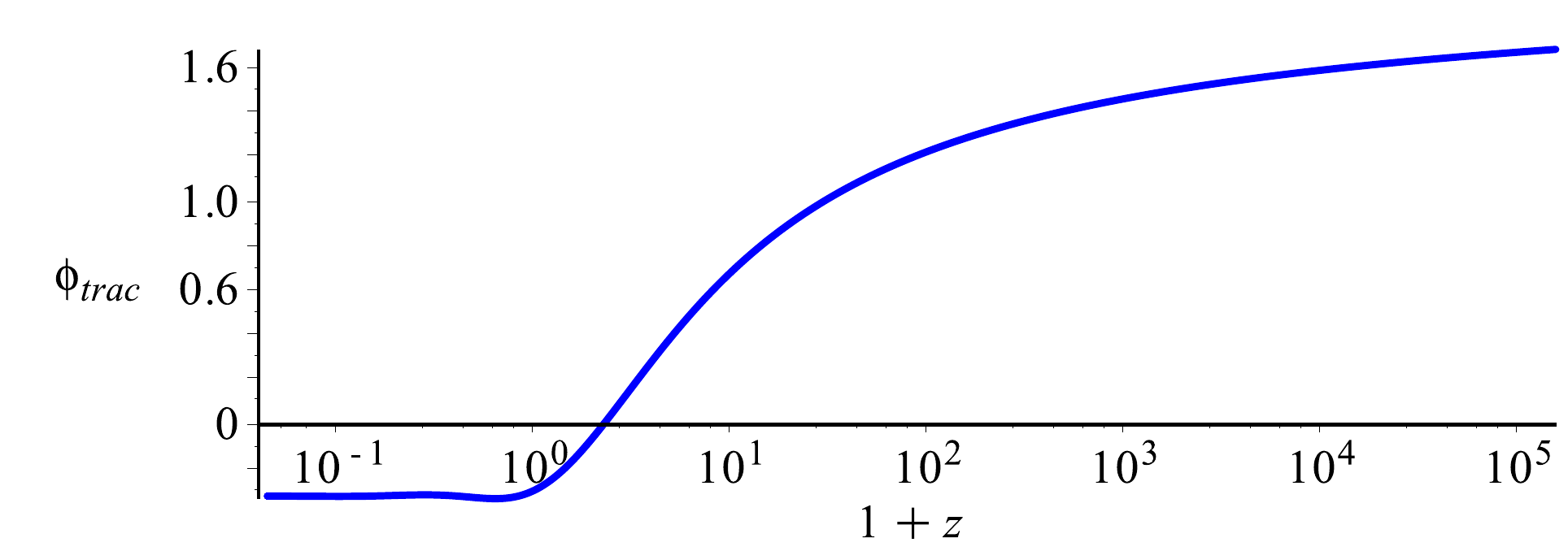,  width=240pt}};
      \draw[->,color=red] (-1.353,9.9) -- (-1.353,8.0) node[midway,left] {\rotatebox{90}{present}};
    \node at (0,5.7) {\epsfig{file=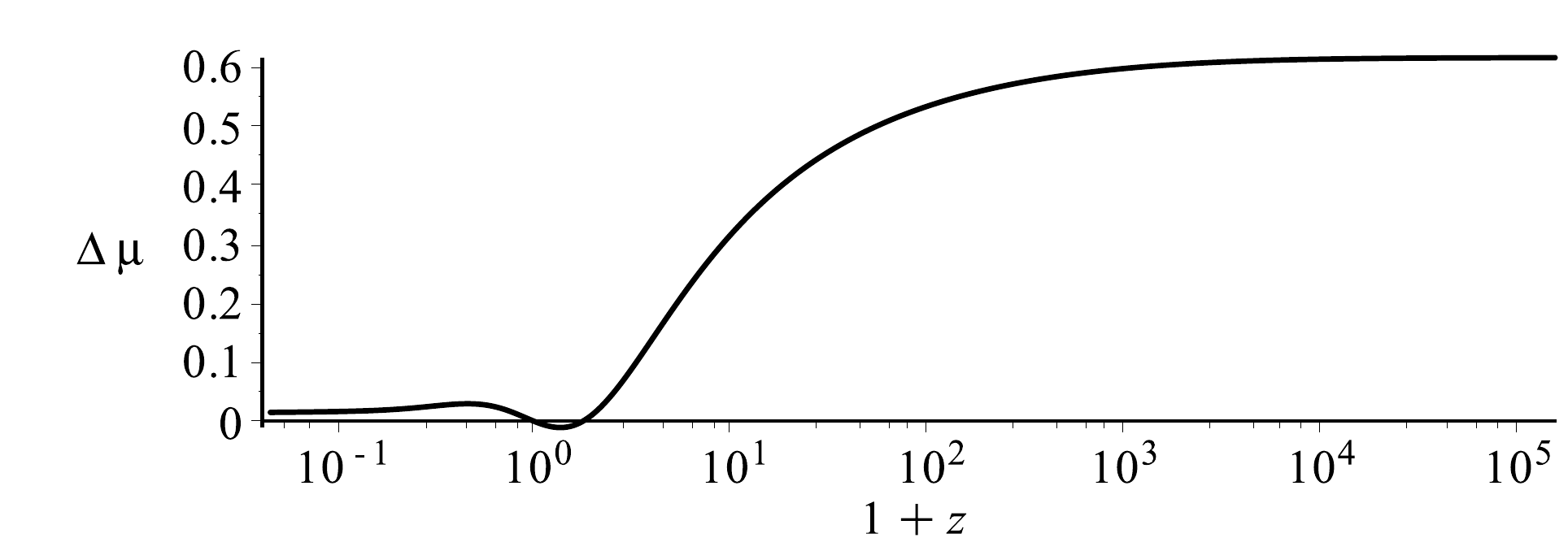, width=240pt}};
    \node at (-1.65,6.3) {\epsfig{file=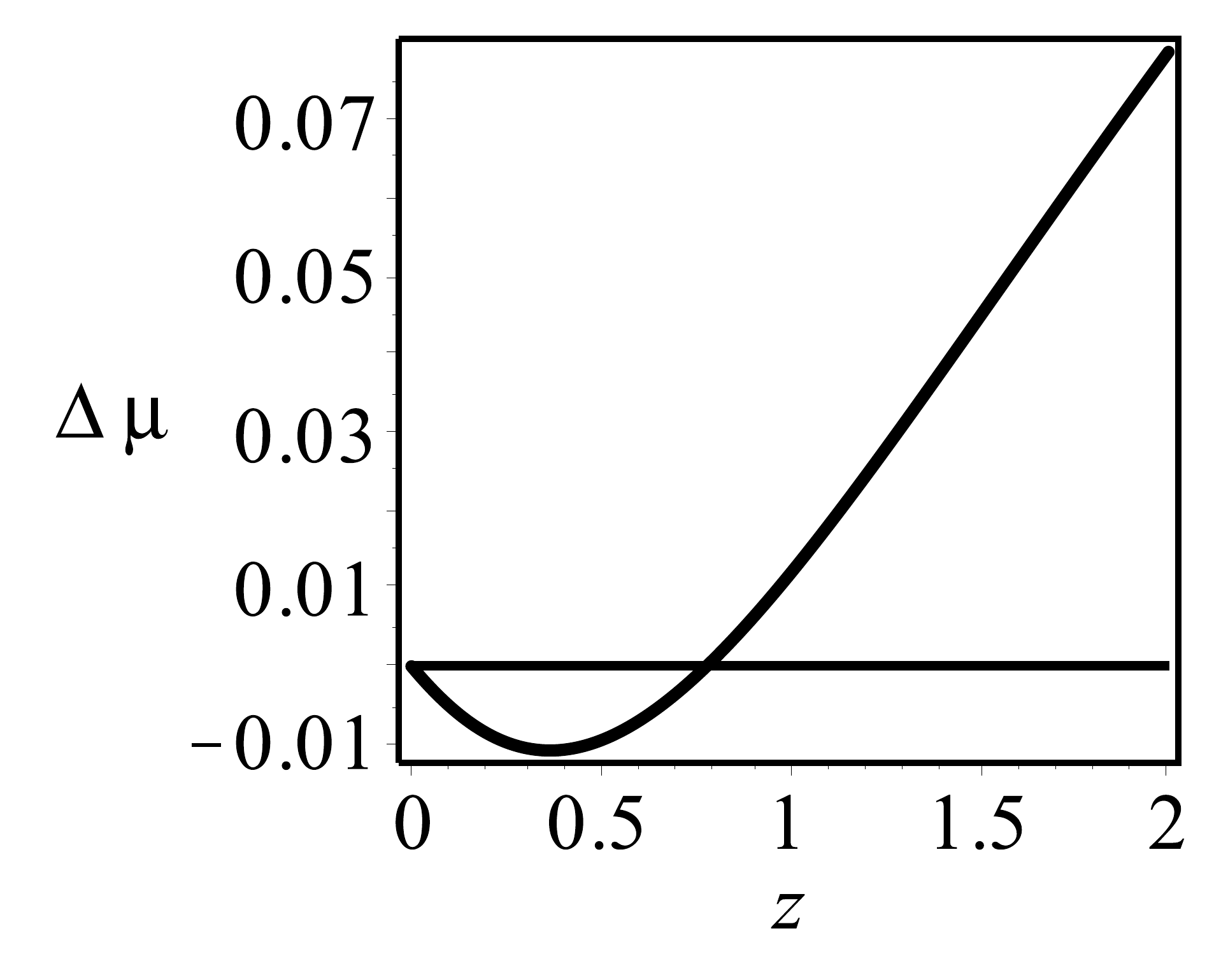, height=52pt}};
      \draw[<-,color=red] (-2.0,5.55) -- (-1.35,5.15);
      \draw[<-,color=red] (-0.7,5.55) -- (-0.86,5.15);
      \draw[color=red] (-1.35,4.85) rectangle (-0.86,5.15);
      \node[above] at (2,5.0) {$\Delta\mu = {}^{+0.08}_{-0.01}$ for $z \in [0,2]$};
    \node at (0,3.0) {\epsfig{file=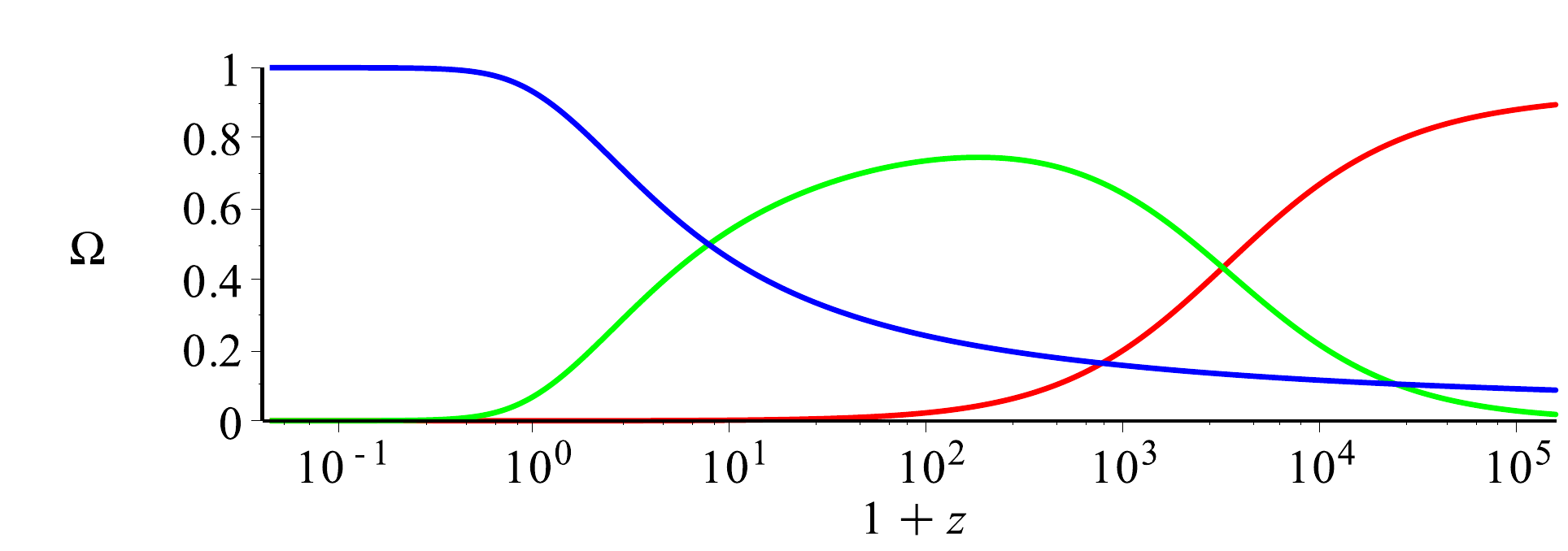, width=240pt}};
      \node[color=red] at (4,3.6) {$\Omega_R$};
      \node[color=green] at (1.8,3.8) {$\Omega_M$};
      \node[color=blue] at (4,2.6) {$\Omega_{\cal Q}$};
    \node at (0,0.0) {\epsfig{file=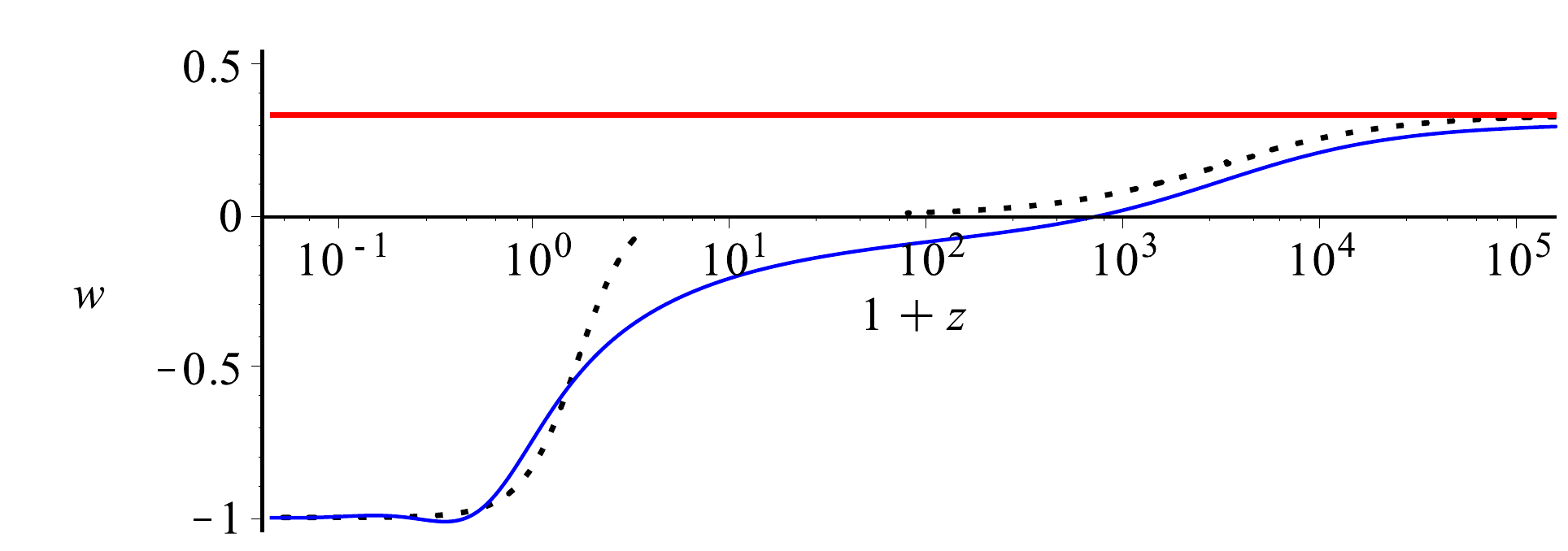, width=240pt}};
      \node[color=blue] at (-2.4,-1.1) {$w_{\cal Q}$};
      \node[color=red] at (3.4,1.1) {$w_{\cal Q} \rightarrow 1/3$};
      \draw[<-] (-0.9,-0.1) -- (0,-0.75) node[right] {approximation $V'(\phi_{\text{eq}}) = {\cal F}$};
  \end{tikzpicture}\\
  (a) \hspace{1em} $\displaystyle f(R) = \left(1 + \ln \frac{R~}{R_0}\right) R$ &
  (b) \hspace{1em} $\displaystyle f(R) = \frac{1 + \ln\frac{R~}{R_0}}{1 + \frac{1}{4}\ln\frac{R~}{R_0}}\, R$\\
  \end{tabular}\end{center}
  \caption{Cosmological evolution of the tracker solutions in the models (\ref{eq:f}) and (\ref{eq:f:2}). Top to bottom: scalaron value $\phi_{\text{trace}}$ for the tracker solution, deviation of the distance modulus $\Delta\mu$ from fiducial $\Lambda$CDM model, effective $\Omega$'s and equation of state $w_{\cal Q}$.}
  \label{fig:modulus}
\end{figure*}
For familiarity's sake, one can introduce effective parameters
\begin{equation}
  \Omega_{\cal Q} = \frac{\rho_{\cal Q}}{3 m_{\text{pl}}^{2} H^2}, \hspace{1em}
  w_{\cal Q} = \frac{p_{\cal Q}}{\rho_{\cal Q}}
\end{equation}
describing scalar field contribution (\ref{eq:Q}) by extracting
\begin{equation}
  \rho_{\cal Q} = 3 m_{\text{pl}}^{2} H^2 - \rho, \hspace{1em}
  p_{\cal Q} = m_{\text{pl}}^{2} (H^2-R/3) - p 
\end{equation}
from evolution history. As mentioned earlier, these are not unique, so for comparison with observations it is much better to directly calculate deviation of distance modulus from fiducial $\Lambda$CDM cosmology
\begin{equation}\label{eq:modulus}
  \Delta\mu = 5\log_{10} \frac{d_{L,\,f(R)}}{d_{L,\,\Lambda\text{CDM}}},\hspace{1em}
  d_L = \frac{1}{a} \int\limits_a^1 \frac{da}{a^2 H}.
\end{equation}
Assuming $\Omega_{M,0} = 0.278$ and matter-radiation equality at redshift $z_{eq}=3250$ \cite{Komatsu:2010fb}, flatness prior fixes $\Omega_{{\cal Q},0} = 0.397$ at present for the tracking solution.
Evolution with redshift is shown in the left panel of Figure~\ref{fig:modulus}. While it is clear that expansion history in our model is substantially different from standard $\Lambda$CDM, deviation of distance modulus up to redshift $z=2$ is only about 1/20 magnitude, which is not constrained by the present SN observations \cite{Riess:2004nr}. Also notable is the fact that effective equation of state of dark energy in this model can appear to cross phantom divide ($w=-1$), and even develop a pole!

\section{Discussion}

The main result of this paper is that running Newton's constant can cause accelerated expansion of the universe. The good news is that the simplest model of this type (\ref{eq:beta:IR},\ref{eq:f}) is predictive and has essentially the same number of parameters as standard $\Lambda$CDM, with exponential hierarchy between effective cosmological constant (\ref{eq:lambda}) and UV scales naturally generated by renormalization group flow (\ref{eq:beta}). The bad news is that while the recent cosmological expansion history is plausibly reproduced, the Newton's constant changes substantially (\ref{eq:eq}) between cosmological, galactic, and near-Earth environments in the simplest realization discussed here. Thus, although the scalar degree of freedom is not light inside matter, this model might have difficulty with gravity tests that probe absolute value of Newton's constant, for example constraints on expansion during nucleosynthesis epoch.

The difficulties with radically changing Newton's constant can be avoided by having a quadratic beta function flow to a finite UV fixed point instead (\ref{eq:beta:2}), which leads to models of the type (\ref{eq:f:2}). These models share similar features with respect to running and hierarchy as the simplest realization we discussed in detail above, but Newton's constant asymptotes to a finite value in the regions of high curvature, as in Einstein's gravity. Taking, for definiteness, $\alpha_* = 1/4$ and $\kappa = 4/3$, leading to
\begin{equation}\label{eq:example}
  f(R) = \frac{1 + \ln\frac{R~}{R_0}}{1 + \frac{1}{4}\ln\frac{R~}{R_0}}\, R,
\end{equation}
one can repeat the analysis of the previous section. While algebraically more complicated, the story goes pretty much along the same lines. Effective potential $V(\phi)$ still has a single minimum and an infinite potential wall at high curvature, but develops a turning point at the low curvature limit. For sensible cosmology, this turning point must be below asymptotic de Sitter curvature in value, which places certain condition on the flow (\ref{eq:beta:2}). The constants in (\ref{eq:example}) are chosen so that this is the case, otherwise no effort was made to tune them to ``special'' values or to fit $\Lambda$CDM fiducial model. With these values, the tracker solution is parametrized by $\Omega_{\cal Q} = 0.722$, and is shown in the right panel of Fig.~\ref{fig:modulus}.

The deviation of distance modulus from fiducial $\Lambda$CDM model is about 1/10 of magnitude out to redshift $z=2$, and $\Omega$ parameters plot looks much more conventional now. The tell-tale sign of this model is much slower dilution of $\Omega_{\cal Q}$ at high redshift than usual, with $\Omega_M$ never quite reaching $1$ during matter domination in this example. This will surely affect the structure formation, and should give a way to test or rule out the model. While parameter space is larger and still remains to be completely explored, this ``soft'' approach to Einstein's gravity is due to logarithmic dependence and is characteristic for all the models we presented here.

Severity of bounds placed by solar system tests \cite{Chiba:2006jp, Hu:2007nk} and large scale structure growth \cite{Pogosian:2007sw, Oyaizu:2008tb, Schmidt:2008tn} will also need to be investigated.

\section*{Acknowledgments}\small
This work was supported by the Natural Sciences and Engineering Research Council of Canada under Discovery Grants program. JQG was supported by the Pacific Century Graduate Scholarship of the Province of British Columbia and the Special Graduate Entrance Scholarship of Simon Fraser University during the academic year 2008/2009. Authors would like to thank Kazuya Koyama, Daniel Litim, Levon Pogosian, Miguel Quartin, Howard Trottier, David Wands, and Kevin Whyte for useful discussions, and Yukawa Institute for Theoretical Physics for hospitality.



\end{document}